\documentclass[floatfix,%
 reprint,
 amsmath,amssymb,
 aps,
]{revtex4-2}

\usepackage{graphicx}% Include figure files
\usepackage{dcolumn}% Align table columns on decimal point
\usepackage{bm}% bold math
\usepackage{xcolor}
\usepackage{soul}
\usepackage{float}

\begin{document}

\preprint{APS/123-QED}

\title{Combined Garvey Kelson Relations for Mass Determinations and Machine Learning}%
\author{I. Bentley}
 \email{ibentley@floridapoly.edu}
\affiliation{Department of Physics, Florida Polytechnic University, Lakeland, FL, 33805}%

\author{A. Fiorito III}
\affiliation{Department of Physics, Florida Polytechnic University, Lakeland, FL, 33805}%

\author{M. Gebran}
\affiliation{Department of Chemistry and Physics, Saint Mary's College, Notre Dame, IN, 46556}%

\author{W. S. Porter}
\affiliation{Department of Physics and Astronomy, University of Notre Dame, Notre Dame, IN, 46556}%

\author{A. Aprahamian}
\affiliation{Department of Physics and Astronomy, University of Notre Dame, Notre Dame, IN, 46556}%

\date{\today}

\begin{abstract}

 Simple Garvey Kelson mass relations applied in two regions are often used as an evaluation metric for machine learning based mass models. These relations have also been used in the training of some machine learning based models. Unfortunately, these Garvey Kelson relations do not broadly sum to zero as is sometimes assumed. In this manuscript, we generate three Garvey Kelson based mass relations that have been optimized with the goal of predicting nuclear masses the most accurately. These three relations have each been optimized for specific tasks. One relation has been optimized to predict the masses on the corner of a 5-by-5 grid. One has been optimized to predict the central mass on that grid, and the last has been optimized to work over the entire grid. Using these relations with the AME 2020 $N \& Z > 7$ data, the central nucleus can be determined with a 129 keV standard deviation, any of the four corner masses with a 472 keV standard deviation, and the overall measure finds a 35 keV standard deviation for a per difference metric. We have compared these results with those from theoretical mass models and have tested the prediction and extrapolation capabilities of the relation that predicts corner masses. We also discuss how these relations can be implemented in machine learning based approaches.

\end{abstract}

\maketitle

\section{Introduction}
Eugene Wigner discussed the use of symmetries among neighboring nuclei to motivate mass relations \cite{PhysRev.51.106}. In this spirit, Garvey Kelson (GK) mass relations consisting of six nuclei have proven to be a successful implementation of symmetry based comparisons. These relations have been derived from an independent particle model, with isospin symmetry where the single particle energies and effective interactions are assumed to vary smoothly \cite{RevModPhys.75.1021}. In these expressions the nucleon-nucleon interactions often do cancel leaving small values remaining \cite{PhysRevLett.16.197,GARVEYG.T.1969SoNR}.

GK relations have been demonstrated to be model independent useful expressions for slowly varying nuclear observables based on physical assumptions \cite{GKSmoothness}. GK relations have been proposed as a means of improving mass models \cite{MoralesIrvingGKtoImprovemasses} and they have also been used to predict other nuclear observables such as charge radii \cite{GKchargeradii}.

The original GK relations, based on mirror symmetry arguments, treated proton rich and neutron rich nuclei differently with separate expressions for the two regions \cite{PhysRevLett.16.197}. Since their inception, it was noted that these expressions result in small but non-zero values and near $N=Z$ this deviation from zero is often the largest \cite{PhysRevLett.16.197}. A third GK expression with a rotated orientation was also introduced for $N=Z$ nuclei which minimizes the discrepancy at $N=Z$ \cite{GARVEYG.T.1969SoNR}. 

In spite of the potential for improvement, the two region form of GK relations have been widely used as an evaluation metric for machine learning based mass models (see e.g., \cite{PhysRevC.109.064322, PhysRevC.111.034305}) as has the three region form in Ref. \cite{DELLEN2024138608}. The two region GK relationships have also been utilized in the training of machine learning models by including them as part of the loss function \cite{PhysRevC.106.L021301}. The two GK relations were also used by Ye et al. as part of an approach to use multiple constraints in model training \cite{PhysRevC.107.044302}. The use of multiple constraints in machine learning based mass models has shown promising results (see e.g., \cite{3x9p-64w7}). A goal of this manuscript is to determine which GK type relations would be preferable over the original expressions for use in model development involving multiple constraints.

There have been efforts to determine which combinations of GK type relations that can be used universally across the chart of the nuclides and that produce values closer to zero. For example, Barea et al. have found that by adding twelve GK relations the standard deviation between the central mass and its predicted value can be reduced \cite{GK20}. This work also demonstrated that it is possible to have one expression that can be applied globally as opposed to using specific relations in different regions. Cheng et al.  have demonstrated that a lower standard deviation can be achieved using some but not all of the twelve GK expressions \cite{PhysRevC.89.061304}. Bao et al. have also found that other mass relations with low standard deviations can be found using some alternative configurations of mass relations \cite{PhysRevC.88.064325}. 

In this manuscript, we discuss how we determined the preferred combinations of GK relations for mass model training and to predict specific unknown masses. Section \ref{MassesSection} contains the definitions of mass evaluation metrics. Section \ref{Mass Differences} discusses GK relations in specific regions, and all possible GK relations that can be generated on a 5-by-5 grid. Section \ref{Methods} describes the approach that was used to determine the optimal combinations. Section \ref{Results} contains the results including how well one of the resulting expressions performs in predicting new masses. Section \ref{discussion} contains a discussion of potential implementation of the results for machine learning based models.

\section{Experimental Masses and Evaluation Metrics}
\label{MassesSection}
One challenge that occurs when attempting to make comparisons among mass relations is that the number of nuclei and therefore the number of nuclei compared, denoted ($N_C$) can vary. To address this we show the metrics with exactly the same set of available nuclei for any given comparison.

Our study will utilize relations in a 5-by-5 grid on the chart of the nuclides centered on a given nucleus. The use of a region of this size is common when dealing with mass relations (see e.g., \cite{GK20,PhysRevC.88.064325,PhysRevC.89.061304}). This size of a region has also been used by Lu et al. to define the space of their convolutional neural network \cite{PhysRevC.111.014325}.

The AME 2020 $N \& Z > 7$ dataset  contains  $N_C$=1365 nuclei for which all of the surrounding nuclei (adding or removing up to two protons and two neutrons) have been experimentally measured  \cite{Wang_2021}. Our determination of which mass relation performs best will be determined using this set of nuclei. 

Another challenge concerns the scale of the comparisons. The evaluation metrics, in this case variations on standard deviations, should produce the same result independent of whether the mass relation is multiplied by some overall scale factor. For this reason we have used two scale free comparisons. 

The standard deviation per difference metric ($\sigma_{PD}$) measures the difference for all of the pairs of nuclei compared. The standard deviation per difference is defined as:
\begin{equation}
    \sigma_{PD}= \sqrt{ \frac{1}{N_C}\sum_{N_C}\bigg(\frac{\Delta M(N,Z)}{N_P}\bigg)^2},
\end{equation}
where $\Delta M(N,Z)$ is the mass relation difference for a specific isotope with ($N$) number of neutrons and ($Z$) number of protons. $N_P$ is the number of differences, which is calculated as half of the sum of the absolute value of all 25 coefficients. The standard GK relations consist of 6 nuclei being compared and the corresponding $N_P$ is three because there are three positive values with three negative values. 

The second metric determines how well the mass difference relation predicts a specific mass at a particular location. The standard deviation from the specific mass ($\sigma_{i}$) is defined as:
\begin{equation}
    \sigma_{i}= \sqrt{ \frac{1}{N_C}\sum_{N_C}\bigg(\frac{\Delta M(N,Z)}{|c_i|}\bigg)^2},
\end{equation}
where $c_i$ is the mass relation coefficient of the specific mass in the difference relation which in principle is divided out as that mass is solved for. In this manuscript we will index the masses in the 5-by-5 grid in reading order from top left to bottom right. We will focus on the four corners: top-left (i=1), top-right (i=5), bottom-left (i=21) and bottom-right (i=25), and the middle (i=13).

\section{Mass Difference Relations}
\label{Mass Differences}
In order for mass difference relations to sum to zero (or nearly zero) they must simultaneously contain the same total number of protons being added and subtracted, and the same total number of neutrons being added or subtracted. Without this constraint the relation will have imbalanced additional mass from the excess protons or neutrons. GK relations are based on symmetries that further constrain the mass relation.

\subsection{Garvey Kelson Relations}
 The original Garvey Kelson relations can be thought of as the sum of three pairs of energy differences. The two relationships from Ref. \cite{GK20}, are:
\begin{equation}
\begin{split}
\label{nGKA9}
    M(N+2,Z-2)-M(N,Z)
    \\+M(N,Z-1)-M(N+1,Z-2)
    \\+M(N+1,Z)-M(N+2,Z-1)\approx 0,
    \end{split}
\end{equation}
for $N\geq Z$, and 
\begin{equation}
    \begin{split}
    \label{GKA1}
    M(N-2,Z+2)-M(N,Z)
    \\+M(N-1,Z)-M(N-2,Z+1)
    \\+M(N,Z+1)-M(N-1,Z+2)\approx 0,
    \end{split}
\end{equation}
for $N<Z$, where $M(N,Z)$ is the mass of an isotope with the corresponding number of neutrons and protons. This will constitute what will be referred to as the two region Garvey Kelson relations. Garvey et al. described the notable differences from zero among $N=Z$ nuclei that these expressions produce \cite{GARVEYG.T.1969SoNR}. 

The updated use of Garvey Kelson relations involves replacing the $N=Z$ values with: 
\begin{equation}
\label{GKB9}
\begin{split}
    M(N+2,Z)-M(N+2,Z-1)
    \\+M(N+1,Z-2)-M(N+1,Z)
    \\+M(N,Z-1)-M(N,Z-2)\approx 0.
    \end{split}
\end{equation}
This will be referred to as the three region GK relations when used in conjunction with the other two. 

\begin{figure*}[t!]
    \centering
    \includegraphics[width=1\linewidth]{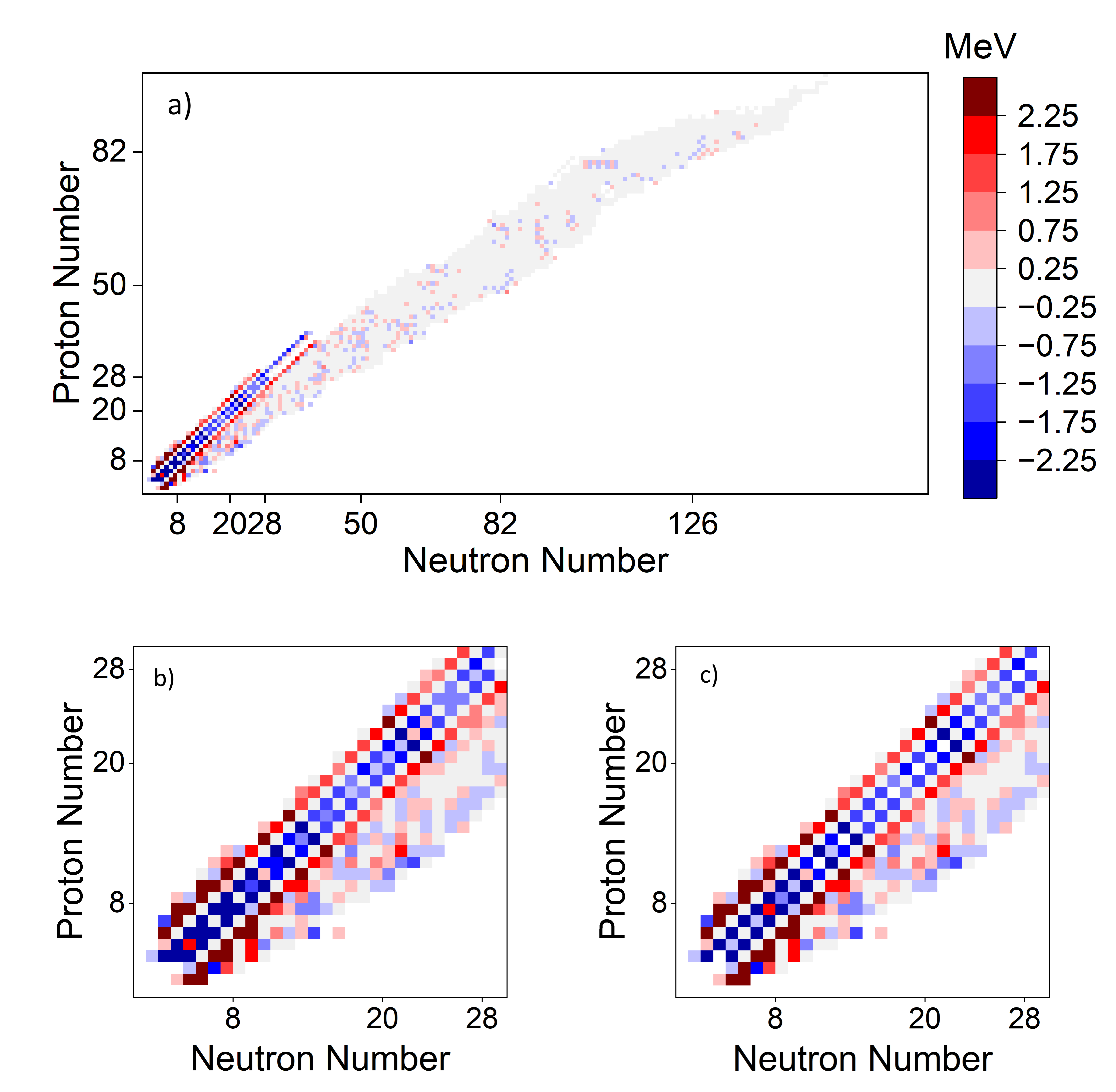}
    \caption{The Garvey Kelson mass relationships for experimental values from AME 2020 \cite{Wang_2021} a) for two regions. A zoomed-in view of the b) two region GK relationships, and c) three region GK relationships in the region around $N=Z$.}
    \label{fig:GK}
\end{figure*}

Figure \ref{fig:GK}a demonstrates the two region GK sum for the experimental nuclear masses from AME 2020 \cite{Wang_2021}. Figures \ref{fig:GK}b and \ref{fig:GK}c have also been included to demonstrate the difference between the two and three region forms and occurs when the third region at $N=Z$ is included. The use of Eqn. (\ref{GKB9}) in the third $N=Z$ region provides a general improvement for the nuclei on that line, but the nearby nuclei continue to have substantial deviations from zero.

\begin{table}[H]
    \caption{Table of GK metric evaluations for the two and three region GK expressions using AME 2020 masses \cite{Wang_2021}.}
    \begin{center}
    \begin{tabular}{|c|cc|cc|}
    \hline
Mass 	&	$N_P$	&	$\sigma_{PD}$	&	$|c_{13}|$		&	$\sigma_{13}$	\\
Relation & & (keV)& &(keV) \\ \hline
2 Region GK	&	3			&	78 &	1	&	234 	\\
3 Region GK	&	3		&	77 &	0 and 1		&	N/A		\\
\hline
    \end{tabular}
    \label{tab1}
      \end{center}
\end{table}

Table \ref{tab1} displays the standard deviation values for these regionally defined GK relationships. It is worth noting that the $\sigma_{13}$ value is not calculable in cases where $c_{13}$=0 because it is not possible to use those relations to predict the mass of the central nucleus. Overall, including the third region does not substantially improve the $\sigma_{PD}$ value.

\subsection{Garvey Kelson Relations on a 5-by-5 grid}

There are two orientations of GK relations  sometimes referred to as transverse and longitudinal (see e.g., \cite{JaneckeTL}) that will be used. These orientations provide the shapes that can be placed at various locations on a 5-by-5 grid. Using these, there are a total of 18 configurations that can be created.

\begin{figure}[H]
    \centering
    \includegraphics[width=1\linewidth]{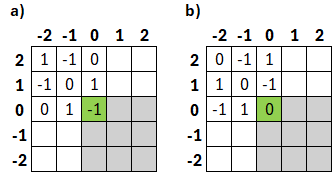}
    \caption{The GK mass model difference configurations displayed as matrices for  a) group A (transverse), b) group B (longitudinal). }
    \label{Fig:GK18}
\end{figure}

Figure \ref{Fig:GK18} demonstrates how the two orientations of GK relations can be cycled through to generate configurations. The groups consist of configurations made by moving the green matrix element to the locations of each of the gray matrix elements. All other non-blank elements are also given the same transform.

In Figure \ref{Fig:GK18}, some of the configurations are related to those used in the two and three region expressions. Specifically,  negative one times configuration A9 corresponds to  Eqn. (\ref{nGKA9}), A1 corresponds to Eqn. (\ref{GKA1}) and B9 corresponds to Eqn. (\ref{GKB9}). 

Table \ref{tab2} includes the $\sigma_{PD}$	and $\sigma_{i}$ values for all 18 GK configurations. The results are generally consistent with one another because they involve many of the same comparisons being used. The differences occur primarily where  configurations have more samples including the ``Wigner Cusp" (discussed in Refs. \cite{SATULA1997103,PhysRevC.66.024327,PhysRevC.88.014322}), and the results of which on GK relations are demonstrated as large magnitude values in Figure \ref{fig:GK}.

Six GK configurations have zero values in the middle and the other twelve have calculable  $\sigma_{13}$ values. There are only four configurations with non-zero values in a corner. The only calculable corner evaluations are $\sigma_{1}$= 0.417 MeV which occurs for A1, $\sigma_{5}$= 0.259 MeV which occurs for B3, $\sigma_{21}$= 0.259 MeV which occurs for B7, and $\sigma_{25}$= 0.267 MeV which occurs for A9.

\begin{table}[H]
    \caption{Table of the $\sigma_{PD}$  and the $\sigma_{13}$ metrics for 18 individual GK configurations in a consistent 5-by-5 grid using the $N_C$=1365 available in AME 2020 \cite{Wang_2021}.}
    \begin{center}
    \begin{tabular}{|c|cc|cc|}
    \hline
Configuration	&	$N_P$	&	$\sigma_{PD}$	&	$|c_{13}|$	&	$\sigma_{13}$		\\
	&		&	(keV)	&		&	(keV)	\\ \hline
A1	&	3	&	139	&	1	&	417	\\
A2	&	3	&	125	&	1	&	376	\\
A3	&	3	&	113	&	0	&	N/A	\\
A4	&	3	&	136	&	1	&	408	\\
A5	&	3	&	119	&	0	&	N/A	\\
A6	&	3	&	103	&	1	&	309	\\
A7	&	3	&	123	&	0	&	N/A	\\
A8	&	3	&	110	&	1	&	331	\\
A9	&	3	&	89	&	1	&	267	\\  \hline
B1	&	3	&	95	&	0	&	N/A	\\
B2	&	3	&	90	&	1	&	271	\\
B3	&	3	&	86	&	1	&	259	\\
B4	&	3	&	94	&	1	&	281	\\
B5	&	3	&	88	&	0	&	N/A	\\
B6	&	3	&	80	&	1	&	241	\\
B7	&	3	&	86	&	1	&	259	\\
B8	&	3	&	80	&	1	&	240	\\
B9	&	3	&	75	&	0	&	N/A	\\  \hline

    \end{tabular}
    \label{tab2}
      \end{center}
\end{table}

\subsection{Other Mass Relations}

The Isobaric Mass Multiplet Equation (IMME) \cite{PhysRev.116.465, WILKINSON1964243, JANECKE1969632} is another mass relation that can be used to predict the mass of one nuclide based on three adjacent  nuclides along an isobaric chain. Limitations have been found with the accuracy of the IMME which have motivated the Quintic Isobaric Mass Multiplet Equation (QIMME) which is an expansion determined using five nuclei along an isobaric chain \cite{PhysRevLett.34.33,PhysRevC.99.014319}. 

Our initial analysis included these mass relations as well. Unfortunately, these relations had various regions where they didn't work well including at $N=Z$ where enhanced proton-neutron pairing at the Wigner Cusp causes a significant deviation from the parabola. We have also tested similar expressions along constant isospin, isotope, and isotone. None of those individual expressions generated a minimum standard deviation below 1 MeV for any metric ($\sigma_{PD}$, $\sigma_{1}$, $\sigma_{5}$, $\sigma_{13}$, $\sigma_{21}$, or $\sigma_{25}$). For this reason we have restricted our approach to only combining GK relations.

\section{Methodology}
\label{Methods}

\subsection{Combined Garvey Kelson Relations}
\begin{table*}[t!]
    \caption{Table of  group combinations for the $\sigma_{PD}$ metric and the $\sigma_{i}$ metrics of interest for the middle and corners of the 5-by-5 grid using the $N_C$=1365 available in AME 2020 \cite{Wang_2021}.}
    \begin{center}
    \begin{tabular}{|c|c|cccccc|}
    \hline
Mass Group	&	GK Relations 	&	$\sigma_{PD}$	&	$\sigma_{1}$	&	$\sigma_{5}$	&	$\sigma_{13}$	&	$\sigma_{21}$	&	$\sigma_{25}$	\\
Label	&	Combined	&	(keV)	&	(keV)	&	(keV)	&	(keV)	&	(keV)	&	(keV)	\\\hline
$\sum M_{12GK}$ from Ref. \cite{GK20}	&	A1-A2-A4+A6+A8-A9-B2+B3+B4-B6-B7+B8	&	60	&	1675	&	1675	&	140	&	1675	&	1675	\\
$\sum M_{8GKPD}$	&	-A6+A9-B1+B3+B4-B6-B7+B8	&	{\bf 35}	&	N/A		&	565	&	282	&	565	&	565	\\
$\sum M_{4GKM}$	&	B3+B4-B6-B7	&	43	&	N/A		&	516	&	{\bf 129}	&	516	&	N/A		\\
$\sum M_{6GKC}$	&	-A1-A5-A9+B1+B5+B9	&	59	&	{\bf 472}	&	{\bf 472}	&	N/A		&	{\bf 472}	&	{\bf 472}	\\\hline
\hline
    \end{tabular}
    \label{tab3}
      \end{center}
\end{table*}

Barea et al., have used the 12 GK Configurations that contain a non-zero value for the central nucleus \cite{GK20}. Here, the GK relations which were rearranged to have a value of negative one at the middle were combined into a single relation consisting of 21 masses of the form:
\begin{equation}
\label{eqn:12GK}
    \begin{split}
    \sum M_{12GK}=\\
            M(N-2,Z+2)+M(N-2,Z-2)\\
            +M(N+2,Z+2)+M(N+2,Z-2)\\
            -2M(N+2,Z-1)-2M(N+2,Z+1)\\
            -2M(N-2,Z-1)-2M(N-2,Z+1)\\
            -2M(N-1,Z-2)-2M(N+1,Z-2)\\
            -2M(N-1,Z+2)-2M(N+1,Z+2)\\
            +2M(N+2,Z)+2M(N-2,Z)\\
            +2M(N,Z-2)+2M(N,Z+2)\\
            +4M(N+1,Z)+4M(N-1,Z)\\
            +4M(N,Z-1)+4M(N,Z+1)\\
            -12M(N,Z)\approx 0,
    \end{split}
\end{equation}
which results in a smaller standard deviation from zero than the standard GK relations do. We will discuss $\sum M_{12GK}$ further when benchmarking our results. 

Unlike the regionally defined GK relations, this expression is ideal because it can be applied universally without accounting for the proximity with the $N=Z$ line. Barea et al. \cite{GK20} and Cheng et al. \cite{PhysRevC.89.061304} both found that combining a subset of GK relations can in some cases result in  lower standard deviations than including all of the expressions will. 

Our methodology extends the approach by including more combinations of GK configurations. Specifically, we did not restrict ourselves to only using configurations that have $|c_{13}|= 1$. Instead, we included all 18 configurations. We have combined them by adding or subtracting each from one another. This method generates $3^{18}=387,420,489$ configurations. It is worth noting that the resulting configurations from this approach included all of the so called generalized GK relationships proposed in Ref. \cite{PhysRevC.88.064325}.

\section{Results}
\label{Results}

The preferred mass relation configurations include the GK combinations for the per difference determination, the middle mass determination, and for the corner mass determination. The GK based mass relation corresponding to the lowest $\sigma_{PD}$ is
\begin{equation}
\label{eqn:GKPD}
    \begin{split}
            \sum M_{8GKPD}=\\
            M(N-1,Z+2) +M(N+2,Z+2)\\
            +M(N+2,Z)+M(N-1,Z-1)\\
            +M(N+2,Z-1)+M(N-2,Z-2)\\
            -M(N,Z+2)-M(N+1,Z+2)\\
            -M(N-2,Z+1)-M(N-1,Z+1)\\
            -M(N+1,Z)-M(N-1,Z-2)\\
            -M(N,Z-2)-M(N+2,Z-2)\\
            +2M(N,Z+1)+2M(N+1,Z+1)\\
            +2M(N-2,Z)+2M(N,Z-1)\\
            +2M(N+1,Z-2)-2M(N+2,Z+1)\\
            -2M(N,Z)-2M(N-2,Z-1)\\
            -2M(N+1,Z-1)\approx 0.
    \end{split}
\end{equation}
The GK based mass relation with the lowest $\sigma_{13}$ is 
\begin{equation}
\label{eqn:GKMiddle}
    \begin{split}
           \sum M_{4GKM}=\\ M(N+2,Z+2)+M(N+1,Z+1)\\
            +M(N-2,Z)+M(N-1,Z)\\
            +M(N+1,Z)+M(N+2,Z)\\
            +M(N-1,Z-1)+M(N-2,Z-2)\\
            -M(N+1,Z+2)-M(N-1,Z+1)\\
            -M(N+1,Z-1)-M(N-1,Z-2)\\
            +2M(N,Z+1)+2M(N,Z-1)\\
            -2M(N+2,Z+1)-2M(N-2,Z-1)\\
            -4M(N,Z)\approx 0.
    \end{split}
\end{equation}
Lastly, the GK based mass relation with the lowest $\sqrt{\sigma_{1}^2+\sigma_{5}^2+\sigma_{21}^2+\sigma_{25}^2}$ value is 
\begin{equation}
\label{eqn:GKC}
    \begin{split}
            \sum M_{6GKC}=\\ 
            M(N-1,Z+2)+M(N+2,Z+2)\\
            +M(N-2,Z+1)+M(N+1,Z+1)\\
            +M(N-2,Z-1)+M(N+1,Z-1)\\
            +M(N-1,Z-2)+M(N+2,Z-2)\\
            -M(N-2,Z+2)-M(N+1,Z+2)\\
            -M(N-1,Z+1)-M(N+2,Z+1)\\
            -M(N-1,Z-1)-M(N+2,Z-1)\\
            -M(N-2,Z-2)-M(N+1,Z-2) \approx 0.
    \end{split}
\end{equation}

\begin{figure*}[t!]
    \centering
    \includegraphics[width=1\linewidth]{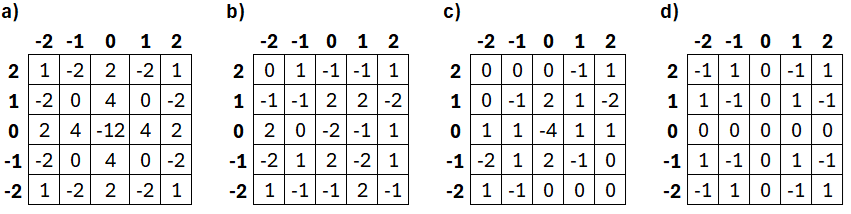}
    \caption{The mass model difference configurations displayed as matrices.  a) Eqn. (\ref{eqn:12GK}) from Ref. \cite{GK20}, b) Eqn. (\ref{eqn:GKPD}), c) Eqn. (\ref{eqn:GKMiddle}), and d) Eqn. (\ref{eqn:GKC}).}
    \label{Fig:best4}
\end{figure*}

Table \ref{tab3} demonstrates the performance of the $\sum M_{12GK}$ and three new relations as well as how each can be generated based on a combination of the original GK configurations shown in Figure \ref{Fig:GK18}. A visual reference of the matrix form of Eqns. (\ref{eqn:12GK}-\ref{eqn:GKC}) has also been provided as Figure \ref{Fig:best4}.

The $\sum M_{8GKPD}$ mass relation results in a $\sigma_{PD}$ value that is less than half of the two and three region GK expressions. The  $\sum M_{4GKM}$ mass relation is more than 100 keV less than the value for the two region relations. Finally, the relation with the lowest corner magnitude could not be calculated using the regional GK expressions. The $\sum M_{6GKC}$ mass relation $\sigma_1$ value is comparable to the $\sigma_1$ value for the A1 GK configuration, but it simultaneously also applies to the three other corners as well.

We further tested more complicated mass relations by  adding the best combinations of configurations to one another and to individual GK configurations. Our determination was that at the cost of substantially higher complexity, with some elements in the hundreds while others were in single digits, the minimum $\sigma_{PD}$ value that can be attained is 33 keV for the experimental data. Similarly, the minimum $\sigma_{13}$ drops to 119 keV with a far more complex mass relation. 

If a simple transform is applied to the 5-by-5 grid of these complex relations, like a mirroring along one axis, or a rotation about the center then the results become less good than those presented in Table \ref{tab3}. For this reason, we have decided to restrict our results to just the expressions shown in Eqns. (\ref{eqn:GKPD}-\ref{eqn:GKC}) involving combinations of the 18 GK relations each used at most once.

\subsection{Testing mass models using the new mass relations}

 Barea et al. have demonstrated that Eqn. (\ref{eqn:12GK}) performs well when applied to theoretical mass models \cite{Massesfarfromstability}. We will similarly test how well theoretical models do with our three mass relations to determine if they are as successful in use with masses from mass models as they were from the experimental data from AME 2020.
 
 The nuclear mass models compared here are the 28 parameter Duflo Zuker (DZ) model \cite{PhysRevC.52.R23}, the Finite-Range Droplet Model (FRDM) 2012 which is based on a liquid droplet with microscopic corrections \cite{MOLLER20161}, the Hartree-Fock-Bogoliubov (HFB) 31 which is based on microscopic interactions and pairing along with phenomenological terms \cite{PhysRevC.93.034337}, and Weizsäcker-Skyrme (WS) 4 \cite{WANG2014215} which is a micro-macro model inspired by Skyrme energy density functionals and a liquid drop model, the WS with Radial Basis Function (WSRBF) correction \cite{PhysRevC.84.051303}, and the Four Model Tree Ensemble (FMTE) which is a composite of four models with machine learning based corrections \cite{PRC112FurtherExploration}.

 Table \ref{tab4} contains the standard deviations for all available masses in each model for a consistent set of $N_C$=5778 nuclei which were available in all models. In this table the metrics are shown for the respective best mass relations from Eqns. (\ref{eqn:GKPD}-\ref{eqn:GKC}). The three evaluation metrics generally demonstrate the smoothness of each model, with low values corresponding to models that are more smooth.
 
  \begin{table}[H]
    \caption{Table of mass relation metrics for the corresponding new mass relations shown in Eqns. (\ref{eqn:GKPD}-\ref{eqn:GKC}) using the common $N_C$ =5778 available in all models.}
    \begin{center}
    \begin{tabular}{|c|ccc|}
    \hline
	&		$\sum M_{8GKPD}$	&	$\sum M_{4GKM}$	&	$\sum M_{6GKC}$	\\ \hline
	&	$\sigma_{PD}$	&	$\sigma_{13}$	&	$\sigma_{1}=\sigma_{5}=\sigma_{21}=\sigma_{25}$	\\
	&		(keV)	&	(keV)	&	(keV)	\\ \hline
DZ \cite{PhysRevC.52.R23}	&		23	&	39	&	275	\\
FRDM \cite{MOLLER20161}	&		35	&	124	&	356	\\
HFB	\cite{PhysRevC.93.034337} &	117	&	430	&	1088	\\
WS \cite{WANG2014215} & 24 & 58 &279\\
WSRBF \cite{PhysRevC.84.051303}	&		27	&	77	&	306	\\
FMTE \cite{PRC112FurtherExploration}	&	29	&	90	&	315	\\ \hline
\hline
    \end{tabular}
    \label{tab4}
      \end{center}
\end{table}

 It is important to note that this metric performance doesn't directly correlate with the success of a model in predicting new masses. Instead it provides an alternative criteria from which behavior similar to that seen in experimental data can be tested. In Table \ref{tab4}, some models performed better regarding these metrics than the experimental data and some performed less well, with the FRDM resulting in values very similar to what was found for the smaller dataset of experimental data. The HFB model stands out as being the least like the experimental data regarding these metrics resulting from an overall less smooth array of masses.

\begin{table*}[t!]
    \caption{Thirteen new precision mass excess measurements and the mass excess deviations, $\Delta ME= ME_{expt.}-ME_{theo.}$, from various models and the prediction using Eqn. (\ref{eqn:GKC}).}
    \begin{center}
    \begin{tabular}{|cccc|cc|ccccccc|}
    \hline
Ref.	&			Element	&	N	&	Z	&	ME	&	Uncertainty	&	$\Delta ME_{DZ}$	&	$\Delta ME_{FRDM}$	&	$\Delta ME_{HFB}$	&	$\Delta ME_{WS}$	&	$\Delta ME_{WSRBF}$	&	$\Delta ME_{FMTE}$	&	$\Delta ME_{GKC}$	\\
&	&				&			&	(keV)	&	(keV)	&	(keV)	&	(keV)	&	(keV)	&	(keV)	&	(keV)	&	(keV)	&	(keV)	\\ \hline
\cite{PhysRevC.104.065803}	&$^{	60	}$	Ga	&	29	&	31	&	-40005	&	30	&	165	&	-1262	&	-965	&	-420	&	-716	&	-109	&	243	\\
\cite{PhysRevC.111.014327}	&$^{	74	}$	Sr	&	36	&	38	&	-40970	&	31	&	-400	&	161	&	540	&	-560	&	6	&	19	&	-413	\\
\cite{s41567-021-01326-9}	&$^{	99	}$	In	&	50	&	49	&	-61429	&	77	&	-479	&	575	&	-209	&	-426	&	21	&	-330	&	-327	\\
\cite{PhysRevLett.127.112501}	&$^{	153	}$	Yb	&	83	&	70	&	-47102	&	46	&	-81	&	200	&	-42	&	57	&	-71	&	-5	&	-140	\\\hline
	
 \cite{SILWAL2022137288}	&$^{	65	}$	Cr	&	41	&	24	&	-28208	&	45	&	-658	&	423	&	382	&	-29	&	146	&	-70	&	-460	\\
\cite{GIRAUD2022137309}	&$^{	74	}$	Ni	&	46	&	28	&	-48451.4	&	3.5	&	9	&	897	&	359	&	767	&	164	&	105	&	-37	\\
\cite{PhysRevC.109.035804}	&$^{	88	}$	As	&	55	&	33	&	-50677.8	&	3.1	&	583	&	-806	&	-28	&	-105	&	-250	&	-57	&	-332	\\
\cite{HUKKANEN2024138916}	&$^{	104	}$	Y	&	65	&	39	&	-53995	&	16	&	-295	&	609	&	-485	&	-334	&	-325	&	-174	&	-264	\\
\cite{HUKKANEN2024138916}	&$^{	106	}$	Zr	&	66	&	40	&	-58582.7	&	4.3	&	948	&	1311	&	-93	&	-162	&	173	&	-34	&	-72	\\
\cite{PhysRevC.103.025811}	&$^{	133	}$	In	&	84	&	49	&	-57678	&	41	&	603	&	542	&	142	&	453	&	332	&	109	&	-198	\\
\cite{PhysRevC.110.045810}	&$^{	152	}$	Ce	&	94	&	58	&	-58891	&	17	&	210	&	595	&	-2001	&	443	&	286	&	1	&	709	\\
\cite{PhysRevC.105.L052802}	&$^{	152	}$	Ce	&	94	&	58	&	-58878.3	&	2.3	&	222	&	607	&	-1988	&	456	&	298	&	14	&	722	\\
\cite{PhysRevC.110.045809}	&$^{	170	}$	Dy	&	104	&	66	&	-53714.6	&	6.3	&	356	&	136	&	-445	&	274	&	255	&	-31	&	82	\\ \hline
	&				&		&		&		&	$\overline{AE}$&	385 &	625	&	591	&	345	&	234	&	81	&	308
	\\\hline
\hline
    \end{tabular}
    \label{tab5}
      \end{center}
\end{table*}

\subsection{Predicting newly measured masses}
One use for GK relations is to predict unknown masses \cite{RevModPhys.75.1021}. Here, we will use the GK Corners mass difference relation. As a first step we analyzed new measurements (since AME 2020) for those where the 24 neighboring nuclei in the 5-by-5 grid had known masses in AME 2020. For this comparison, we restricted our analysis to precision measurements with experimental uncertainties under 100 keV.

Table \ref{tab5} demonstrates the Mass Excess ($ME$) values for thirteen new measurements that fit these criteria from Refs. \cite{PhysRevC.104.065803,PhysRevC.111.014327,s41567-021-01326-9,PhysRevLett.127.112501,SILWAL2022137288,GIRAUD2022137309,PhysRevC.109.035804,HUKKANEN2024138916,PhysRevC.103.025811,PhysRevC.110.045810,PhysRevC.105.L052802,PhysRevC.110.045809}. Four of these measurements,  nearer the proton rich side, had a mass in the top-left location that was solved for assuming the relation shown in Eqn. (\ref{eqn:GKC}) summed to zero. Similarly, the later nine neutron rich masses were determined using the same equation by solving for the one unknown mass in the bottom-right corner.

The newly measured mass excess values are compared with  theoretical models from Refs. \cite{PhysRevC.52.R23,MOLLER20161,PhysRevC.93.034337,WANG2014215,PhysRevC.84.051303,PRC112FurtherExploration}. The comparison with predictions involving Eqn. (\ref{eqn:GKC}) are included in the last column of the table. The mean Absolute Error ($\overline{AE}=\sum |\Delta ME|/N_C$) for the thirteen $ME$ values compared to the theoretical predictions are provided in the last row of Table \ref{tab5}. In this comparison, only the FMTE model and WSRBF both perform better than the $\sum M_{6GKC}$ at predicting these new masses. It is worth noting that the standard deviation was 384 keV for the 13 predictions using $\sum M_{6GKC}$. That is about 90 keV lower than its performance with the full AME 2020 indicating that the predictions using these relationships can outperform the values listed in Table \ref{tab3}.

\subsection {Extrapolations based on the corners}
Morales et al. have demonstrated how an iterative process can be used to predict masses further from stability  \cite{MORALES2009113}. We have similarly used Eqn. (\ref{eqn:GKC}) to predict missing corner masses on a 5-by-5 grid for neutron rich nuclei. 

Our iterative process involved two steps. Step 1 is to generate all possible masses where the neighboring 24 masses are known (or have been solved for) to determine the bottom-left mass on the 5-by-5 grid. Step 2 is to similarly generate the bottom-right mass where the neighboring 24 masses are known or have been solved for. Then repeat step 1 and step 2, to fill in the remaining nuclei. 

We also tested the same process but beginning with the order reversed (bottom-right followed by the bottom-left, and then repeated), to test the sensitivity on the order. Across the chart of the nuclides we found that 78\% of the extrapolated masses were order independent and the largest deviations between the two approaches were less than 8 MeV.

\begin{figure}[t!]
    \centering
    \includegraphics[width=1.0\linewidth]{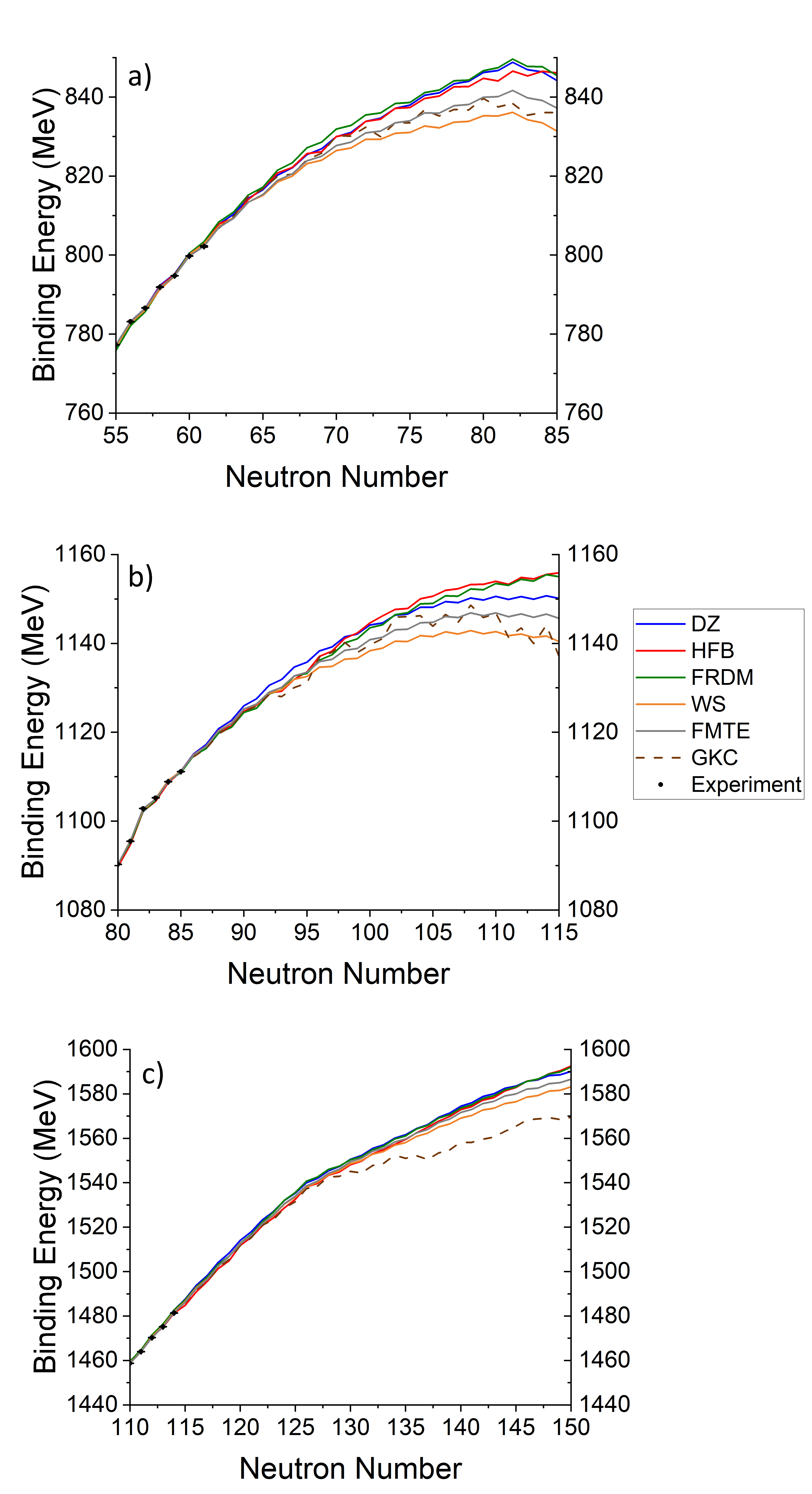}
    \caption{Mass model extrapolation comparison for neutron-rich (a) krypton ($Z = 36$), (b) tin ($Z = 50$), and (c) hafnium ($Z = 72$) isotopes. Experimental values from AME 2020 \cite{Wang_2021} are included as black circles. Solid lines indicate the existing mass models from Refs. \cite{PhysRevC.52.R23,PhysRevC.93.034337,MOLLER20161,WANG2014215,PRC112FurtherExploration}. The brown dashed line shows the mass prediction using an iterative procedure involving  solving for the corner mass based on Eqn. (\ref{eqn:GKC}). }
    \label{fig:parab}
\end{figure}

Figure \ref{fig:parab} shows the extrapolation of neutron rich masses using various mass models. These are available as reference against the average of the two iterative processes. The use of $\sum M_{6GKC}$ generally predicts masses that are consistent with the other mass models for the first ten predictions of neutron rich nuclei past what is known experimentally. For predictions further from stability the deviations become greater. Figure \ref{fig:parab}c demonstrates that for neutron rich hafnium isotopes near $N=150$, the five models are all within about 10 MeV of each other. The use of $\sum M_{6GKC}$ to reach 35 additional neutrons from the last known experimental value results in a binding energy value that is about 10 MeV below the lowest value in any of the models.

\section{Discussion for implementation in machine learning based mass models}
\label{discussion}

The combined GK relations shown in Eqns. (\ref{eqn:GKPD})-(\ref{eqn:GKC}) provide not only an evaluation metric for mass models, but also a natural regularization framework for machine learning based mass predictions. Rather than training a machine learning model to minimize prediction mass residual alone, we propose embedding the GK relations directly into the loss function, constraining the model to produce a mass surface that is consistent with the smoothness properties observed in experimental nuclear masses. While this concept has been incorporated into machine learning approaches previously \cite{PhysRevC.106.L021301}, the use of more appropriate, refined GK relations, as presented here, could significantly improve the accuracy and smoothness of a given model.

This process can be applied to data including one or multiple mass models. If multiple mass models are used this approach could include a multi-target extension where multiple models are simultaneously optimized together.

One could envision also including the corresponding metric values from the experimental data, shown in Table \ref{tab3} as the target for the machine learning model. Overall, this approach can provide a unified, physically constrained correction framework applicable to any of the mass models considered. It can also be extended to include new mass models as they become available.

\section{Summary and Conclusion}
One purpose of this work was to demonstrate that expecting the two and three region GK relations to be zero for mass model evaluation is flawed because these expressions do not sum to zero, particularly near $N=Z$. Instead the use of combined GK relations can result in expressions that are both closer to zero and they can be applied across the chart of the nuclides.

Building primarily on the work of combining GK relations from Refs. \cite{GK20,PhysRevC.89.061304,PhysRevC.88.064325}, we have created three new mass relations that are optimized for individual constraints and that have no regional dependence. These utilize a 5-by-5 grid of neighboring masses to produce minimal mass differences for:  \begin{itemize}
    \item Eqn. (\ref{eqn:GKPD}) the pairs of nuclei compared,
    \item Eqn. (\ref{eqn:GKMiddle}) predictions of the central mass in the middle of the grid, and
    \item Eqn. (\ref{eqn:GKC}) predictions of a mass at a corner of the grid.
\end{itemize}

It is important to note that the three newly determined mass relations, like all prior GK relations, don't sum to zero as the result of underlying shell structure (see e.g., \cite{MORALES2009113,PhysRevC.93.044337,RevModPhys.92.015002}). With that stated, these relations provide a preferable alternative for specific uses. Any one of them, or perhaps all three in tandem, can be used to validate mass models. Eqn. (\ref{eqn:GKC}) can be used to determine missing masses at the boundary of what is known. 

We have tested how consistent theoretical mass models are compared to the experimental data using these expressions. Eqn. (\ref{eqn:GKC}) was also used to solve for 13 newly measured masses. The corresponding level of success was generally competitive with state-of-the-art mass models. This same mass relation was also used to predict missing masses far from stability as part of an iterative process. Here the neutron rich predictions are consistent with other theoretical models for the first ten predictions with less agreement further from stability.

The mass relations corresponding to Eqns. (\ref{eqn:GKPD}-\ref{eqn:GKC}) are promising for use in training and testing machine learning based mass models. This could come in the form of incorporating one or more of them as a training feature, as part of the loss function, or to motivate physical constraints on convolutional neural networks that use nearby neighbors as inputs.

\bibliography{bibliography}

\end{document}